\documentstyle[12pt]{article}
\makeatletter
\ifcase \@ptsize
\or 
\or 
\fi
\advance\textheight by \topskip

\ifcase \@ptsize
\or 
\or 
\fi

\def\ga{\gamma}         
\def\be{\beta}
\def\al{\alpha}
\def\ep{\epsilon}
\def\la{\lambda}        
\def\de{\delta}         
         
\def\sig{\sigma}        
      \def\vth{\vartheta}

\def\an{$a_n^{(1)}$}
\def\d4{$d_4^{(1)}$}
\def\cn{$c_n^{(1)}$}
\def\dn{$d_n^{(1)}$}
\def\bn{$b_n^{(1)}$}
\def\esix{$e_6^{(1)}$}
\def\esev{$e_7^{(1)}$}
\def\eigh{$e_8^{(1)}$}
\def\gtwo{$g_2^{(1)}$}
\def\hir{(D_t^2-D_x^2)}

\begin{document}
\renewcommand{\thefootnote}{\fnsymbol{footnote}}
\baselineskip 20pt

{\pagestyle{empty}
\rightline {DTP-92-45 / RIMS-890}
\rightline {July 1992}
\vskip 1in
\centerline{\LARGE Affine Toda solitons and automorphisms of Dynkin diagrams}
\vskip 0.5in
\centerline {\it \Large Niall J. MacKay\footnote{mackay@kurims.kyoto-u.ac.jp}}
\centerline {\it Research Institute for Mathematical Sciences,}
\centerline {\it University of Kyoto,}
\centerline {\it Kyoto 606,}
\centerline {\it Japan.}
\vskip 0.20in
\centerline {\it \Large William A. McGhee\footnote{W.A.McGhee@durham.ac.uk}}
\centerline {\it Department of Mathematical Sciences,}
\centerline {\it University of Durham,}
\centerline {\it Durham DH1 3LE,}
\centerline {\it England.}
\medbreak
\vskip 0.8in
\centerline {\Large Abstract}
\vskip 1cm
Using Hirota's method, solitons are constructed for affine Toda field
theories based on the simply-laced affine algebras. By considering
automorphisms of the simply-laced Dynkin diagrams, solutions to the
remaining algebras, twisted as well as untwisted, are deduced.
\vfill
\newpage}
%
%
\section{Introduction}

\bigbreak
\pagenumbering{arabic}
Recent work has shown that soliton solutions can be constructed for
affine Toda field theories based on the \an\ and \d4\ algebras \cite{holl}
 as
well as the \cn\  algebra \cite{sao} when the coupling constant is purely
complex.
 In the case of the \an\ theory N-soliton solutions have been
constructed, whereas for \d4\ and \cn\ only static single solitons.
\medbreak
The purpose of this paper is to construct static single solitons for
all of the remaining algebras, both twisted and untwisted. This is
achieved by considering a generalisation of the field ansatz used in
\cite{holl}, although as in \cite{holl} a special decoupling of the equations
of
motion is considered. It is found that once the soliton solutions for
\dn,\ \esix,\ \esev \ and \eigh\ are constructed, solutions for
theories based on the other algebras follow by folding the simply-laced
Dynkin diagrams. The solutions for \esix and \esev have been obtained
independently by Hall \cite{rah}.
\medbreak
For the simply-laced algebras the number of static solitons is found
to be equal to the rank of the corresponding algebra. However by the
method that will be employed here, for the non-simply-laced algebras
a lesser number is found (as in \cite{sao} for \cn\ ). Also, for all the
theories
the mass ratios of the solitons\footnote{For \gtwo\  and \cn\ only one
soliton is found and so mass ratios cannot be considered.} can be
calculated and are found to coincide with the mass ratios of the
fundamental particles in the real-coupling affine Toda theory ({\em
i.e.\ }those obtained by expanding
the potential term of the Lagrangian density about its minimum \cite{mop}
\cite{bcds1}).
\medbreak
The paper concludes with a discussion of some aspects of topological charge.
\bigbreak
%
%
\section{The equations of motion}

\bigbreak
The Lagrangian density of affine Toda field theory can be written in the form

$${\cal L}={1\over2}(\partial_\mu \phi)(\partial^\mu\phi)-{m^2\over\be ^2}\sum_
  {j=0}^n n_j(e^{\be\al_j\cdot\phi}-1).$$

The field $\phi(x,t)$ is an $n$-dimensional vector, $n$ being the rank
of the finite Lie algebra $g$. The $\alpha_j$'s, for $j=1,...,n$ are
the simple roots of $g$;
$\alpha_0$ is chosen such that the inner products among the elements
of the set
\{$\al_0,\al_j$\} are described by one of the extended Dynkin
diagrams. It is expressible in terms of the other roots by the equation

$$\al_0=-\sum_{j=1}^n n_j\alpha_j$$

\noindent where the $n_j$'s are positive integers, and $n_0=1$.
Both $\be$ and m are constants, $\be$ being the coupling constant.
\medbreak
The inclusion of $\al_0$ distinguishes affine Toda field theory from
Toda field theory. Toda field theory is conformal and integrable, its
integrability implying the existence of a Lax pair, infinitely many
conserved quantities and exact solubility \cite{mop}\cite{wilson}\cite{ot}
(for further references see \cite{lez}).
The extended root
is chosen in such a way as to preserve the integrability of Toda field
theory (though not the conformal property), with the enlarged set of
roots $\{\al_0,\al_j\}$ forming an admissible root system \cite{mop}.
\medbreak
Setting the coupling constant $\be$ to be purely complex, {\em i.e.\
}$\be =i\ga$,
the equations of motion are

$$\partial ^2 \phi - {im^2\over\ga}\sum_{j=0}^n n_j \alpha_j e^{i\ga
\alpha_j\cdot\phi}=0.\eqno (2.1)$$

Extending the idea of \cite{holl}, consider the following substitution for the
field $\phi(x,t)$\footnote{For the simply-laced algebras the choice of
$\eta_i$ coincides with that of \cite{holl}, namely \mbox{$\eta_i=1$.}
However, for the remaining algebras it is other choices of $\eta_i$
which yield soliton solutions.},

$$\phi=-{1\over i\ga}\sum_{i=0}^n \eta_i\alpha_i \ln\tau_i$$

\noindent which reduces (2.1) to the form

$$\sum_{j=0}^n\alpha_j Q_j=0$$

\noindent where

$$ Q_j = \left ({\eta_j\over \tau_j^2} (D_t^2-D_x^2)\tau_j\cdot
\tau_j-2m^2n_j \left (\prod_{k=0}^n \tau_k ^{-\eta_k\alpha_k\cdot\alpha_j}-1
\right )\right ).$$

\noindent $D_x$ and $D_t$ are Hirota derivatives, defined by
\medbreak
$$D_x^m D_t^n f\cdot g=\left ({\partial \over \partial x}- {\partial
\over \partial x^{\prime}} \right ) ^m \left ({\partial \over \partial
t}- {\partial \over \partial t^{\prime}} \right ) ^n
f(x,t)g(x^{\prime},t^{\prime}) \left \vert \right . _{x=x^{\prime}
\atop t=t^{\prime}}.$$
\medbreak
It will be assumed ({\em cf} \cite{holl}) that $Q_j=0 \ \forall j$, although
this is not the
most general decoupling. (The existence of $n+1$ $\tau$-functions
(compared to the $n$-component field $\phi$) is due to the
relationship between affine
and conformal affine Toda theories \cite{sao}.)  Therefore,
\medbreak
$$\eta_j (D_t^2-D_x^2)\tau_j\cdot\tau_j-2m^2 n_j \left ( \prod_{k=0}^n
\tau_k^{-\eta_k\alpha_k\cdot\alpha_j}-1 \right )\tau_j^2=0. \eqno(2.2)$$

In the spirit of Hirota's method for finding soliton solutions \cite{hir},
suppose

$$\tau_j=1+\de_j^{(1)}e^\Phi \ep+\de_j^{(2)}e^{2\Phi} \ep^2+ .... +
\de_j^{(p_j)}e^{p_j\Phi} \ep^{p_j}$$

\noindent where $\Phi=\sig (x-vt-\xi)$ and $\de_j^{(k)} (1\leq k \leq p_j),
\sig , v$ and $\xi$ are arbitrary complex constants. The constant
$p_j$ is a positive integer and $\ep$ an infinitesimal parameter. The
method employed is to solve (2.2) at successive orders in $\ep$, and
then absorb $\ep$ into the exponential.
\medbreak
At first order in $\ep$, it is easily shown that
\medbreak
$$\sum_{j=0}^n K_{ij} \delta_j^{(1)} = {\sigma^2(1-v^2) \over m^2}
\delta_i^{(1)} $$
\noindent where
$$ K_{ij}={\eta_j \over \eta_i}n_i \alpha_i\cdot\alpha_j.$$
\medbreak
\noindent Defining the matrices,
\begin{enumerate}
\item $\eta=\hbox{diag}(\eta_0,\eta_1,...,\eta_n)$
\item $N=\hbox{diag}(n_0,n_1,...,n_n)$
\item $(C)_{ij}=\alpha_i\cdot\alpha_j$
\end{enumerate}
\medbreak
\noindent then
$\delta^{(1)}=(\delta_0^{(1)},\delta_1^{(1)},...,\delta_n^{(1)})^{T}$
is an eigenvector of the matrix $K$ where
\medbreak
$$\eta K \eta^{-1}=NC$$
\medbreak
\noindent with eigenvalue $\lambda$ where
\medbreak
$$\sigma^2 (1-v^2)=m^2 \lambda.$$
\medbreak
\noindent As $K$ and $NC$ are similar, they share the same
eigenvalues. Indeed for the a,d and e theories it has been shown \cite{bcds}
that
the squared masses of
the fundamental Toda particles are eigenvalues of $NC$. For the
non-simply-laced
theories, the eigenvalues of $NC$ are also eigenvalues of a
simply-laced theory and so are related to the squared masses of the
non-simply-laced theory. As will be
seen in section 5 this leads to the ratios of static energies of the
solitons being equal to the ratios of the unrenormalized masses of the
fundamental particles described by the Lagrangian fields.
\medbreak
It is straightforward to show that for $\tau_j$ to be bounded as $x\rightarrow
\pm\infty$,

$$n_0\eta_j p_j = n_j\eta_0 p_0.$$

\noindent In all cases, $\eta_j$ is chosen to be
$$\eta_j={2\over \al_j\cdot\al_j},$$
\noindent since this choice of $\eta_j$ causes each $\tau_j$ to be
raised to a non-negative integer power in the equations of motion
(2.2). So, for the simply-laced cases \mbox{$\eta_j=1$} and for
single soliton solutions $p_j=n_j$.
\medbreak
Finally, it is unnecessary to consider the solution corresponding to
$\la=0$, as it is always $\phi=0$.
\bigbreak
%
%
\section{Affine Toda solitons for simply-laced algebras}
\bigbreak
The length of the longest roots will be taken to be $\sqrt 2$ for all
cases. It is necessary fix the root lengths in this way, otherwise the
parameters $m$ and $\beta$ in the equations of motion have to be
rescaled.
Also under this convention, the soliton masses are found to satisfy
one universal formula.

%
%
\vfill
\newpage
\subsection{The \an\ theory}
\bigbreak
The Dynkin diagram for \an\ is shown in Figure 3.1a.
\medbreak
The eigenvalues of the matrix $NC$ are
$$\la_a=4\sin^2\left ({\pi a\over n+1}\right ).$$
\bigbreak
\
\bigbreak
\centerline{
\begin{picture}(250,125)(0,-20)
\put ( 18,50){\line( 1, 0){24}}
\put ( 58,50){\line( 1, 0){24}}
\put ( 98,50){\line( 1, 0){14}}
\put (168,50){\line( 1, 0){24}}
\put (208,50){\line( 1, 0){24}}
\put (152,50){\line(-1, 0){14}}
\put (132,95){\line(5,-2){102}}
\put (118,95){\line(-5,-2){102}}
\put ( 10,50){\circle{10}}
\put ( 50,50){\circle{10}}
\put ( 90,50){\circle{10}}
\put (160,50){\circle{10}}
\put (200,50){\circle{10}}
\put (240,50){\circle{10}}
\put (125,98){\circle{10}}
\put (10,35){\makebox(0,0){$\alpha_1$}}
\put (50,35){\makebox(0,0){$\alpha_2$}}
\put (90,35){\makebox(0,0){$\alpha_3$}}
\put (160,35){\makebox(0,0){$\alpha_{n-2}$}}
\put (200,35){\makebox(0,0){$\alpha_{n-1}$}}
\put (240,35){\makebox(0,0){$\alpha_n$}}
\put (125,114){\makebox(0,0){$\alpha_0$}}
\put (125,50){\makebox(0,0){.....}}
\put (125,10) {\makebox(0,0){Figure 3.1a: Affine Dynkin diagram for \an.}}
\end{picture}
}
With $\eta_j=1 \ \forall j$, the equations of motion are

$$(D_t^2-D_x^2)\tau_j\cdot\tau_j=2m^2 (\tau_{j-1} \tau_{j+1}-\tau_j^2)$$

\noindent {\em i.e.\ }those of \cite{holl}. For the single soliton solutions
$p_0=1$, giving
$$\tau_j=1+\omega^j e^{\Phi}$$
\noindent where $\omega$ is an (n+1)$^{\hbox{th}}$ root of unity.
There are $n$ non-trivial solutions \cite{holl} (equal to the number of
fundamental particles) with $\omega_a=\hbox{exp 2}\pi ia/\hbox{(n+1)}$
where
$1\leq a\leq n$.
These $n$ solutions to \an\ can be written in the form
\medbreak
$$\phi_{(a)}=-{1\over i\ga}\sum_{k=1}^r \alpha_j\ln\left ({1+w_a^j
e^{\Phi} \over 1+e^{\Phi}}\right ).$$
\medbreak
It was shown in \cite{holl} that $\phi_{(a)}$ $(1\leq a\leq n)$ can be
associated with the $a$-th fundamental representation of \an, and that
different values of $Im\,\xi$ give rise to different topological
charges. The topological charges are found to be weights of the
particular representation. Therefore, strictly speaking the results
presented here correspond to representatives from each class of
solution, as the value of $\xi$ and so the topological charge, is not
specified.
\medbreak
\subsection{The \dn\ theory}
\bigbreak
The equations of motion for $d_4^{(1)}$, whose Dynkin diagram is shown
in Figure 3.2a, are slightly different to those for $d_{n\geq
5}^{(1)}$\  and so will be considered separately.
\bigbreak
\centerline{
\begin{picture}(100,120)(0,5)
\put ( 45, 85){\line(-1,-1){20}}
\put ( 55, 85){\line( 1,-1){20}}
\put ( 55, 95){\line( 1, 1){20}}
\put ( 45, 95){\line(-1, 1){20}}
\put ( 20, 60){\circle{10}}
\put ( 50, 90){\circle{10}}
\put ( 80, 60){\circle{10}}
\put ( 20,120){\circle{10}}
\put ( 80,120){\circle{10}}
\put ( 20,105){\makebox(0,0){$\alpha_0$}}
\put ( 20, 45){\makebox(0,0){$\alpha_1$}}
\put ( 50, 75){\makebox(0,0){$\alpha_2$}}
\put ( 80,105){\makebox(0,0){$\alpha_3$}}
\put ( 80, 45){\makebox(0,0){$\alpha_4$}}
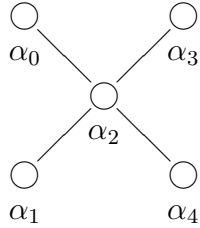
\put ( 50, 20){\makebox(0,0){Figure 3.2a: Affine Dynkin diagram for
$d_4^{(1)}$.}}
\end{picture}
}
The eigenvalues of the matrix $NC$ are $\la=2,2,2$ and $6$. With
$\eta_j=1 \ \forall j$, the single soliton has $p_j=n_j\ \forall j$
and satisfies
\begin{eqnarray*}
(D_t^2-D_x^2)(\tau_j\cdot\tau_j)\!\!\!\! &=& \!\!\!\! 2m^2
(\tau_2-\tau_j^2)\qquad (j\neq 2) \\
(D_t^2-D_x^2)(\tau_2\cdot\tau_2)\!\!\!\! &=& \!\!\!\! 4m^2
(\tau_0\tau_1\tau_3\tau_4-\tau_2^2).
\end{eqnarray*}
\medbreak
If $\la$=2, three solutions are obtained \cite{holl}:
\begin{eqnarray*}
\tau_0\!\!\!\! &=&\!\!\!\!\tau_3=1+e^{\Phi} \\
\tau_2\!\!\!\! &=&\!\!\!\! 1+e^{2\Phi} \\
\tau_1\!\!\!\! &=&\!\!\!\!\tau_4=1-e^{\Phi}
\end{eqnarray*}

\noindent and cycles of the indices (1,3,4).
\medbreak
If $\la$=6, one solution is obtained:
\begin{eqnarray*}
\tau_0 \!\!\!\!&=&\!\!\!\!\tau_1=\tau_3=\tau_4=1+e^{\Phi} \\
\tau_2 \!\!\!\!&=&\!\!\!\! 1-4e^{\Phi}+e^{2\Phi}.
\end{eqnarray*}
\medbreak
The Dynkin diagram for \dn\ $(n\geq 5)$ is shown in Figure 3.2b.
\bigbreak
\centerline{
\begin{picture}(250,120)(0,5)
\put ( 45, 85){\line(-1,-1){20}}
\put ( 58, 90){\line( 1, 0){24}}
\put ( 98, 90){\line( 1, 0){14}}
\put (168, 90){\line( 1, 0){24}}
\put (205, 85){\line( 1,-1){20}}
\put (152, 90){\line(-1, 0){14}}
\put (205, 95){\line( 1, 1){20}}
\put ( 45, 95){\line(-1, 1){20}}
\put ( 20, 60){\circle{10}}
\put ( 50, 90){\circle{10}}
\put ( 90, 90){\circle{10}}
\put (160, 90){\circle{10}}
\put (200, 90){\circle{10}}
\put (230, 60){\circle{10}}
\put ( 20,120){\circle{10}}
\put (230,120){\circle{10}}
\put ( 20,105){\makebox(0,0){$\alpha_0$}}
\put (238,105){\makebox(0,0){$\alpha_{n-1}$}}
\put ( 20, 45){\makebox(0,0){$\alpha_1$}}
\put ( 50, 75){\makebox(0,0){$\alpha_2$}}
\put ( 90, 75){\makebox(0,0){$\alpha_3$}}
\put (160, 75){\makebox(0,0){$\alpha_{n-3}$}}
\put (200, 75){\makebox(0,0){$\alpha_{n-2}$}}
\put (230, 45){\makebox(0,0){$\alpha_n$}}
\put (125, 90){\makebox(0,0){.....}}
\put (125, 20){\makebox(0,0){Figure 3.2b: Affine Dynkin diagram for \dn.}}
\end{picture}
}
In this case the eigenvalues of the matrix $NC$ are
$$\la_a=8\sin^2 \vth_a\ \ \hbox{where} \ \ \vth_a={a\pi\over 2(n-1)}\
\ (1\leq a\leq n-2)$$
$$\hbox{and} \ \ \la_{n-1}=\la_n=2.$$
With $\eta_j=1 \ \forall j$, the single soliton has $p_j=n_j\ \forall
j$ and satisfies the following equations
\begin{eqnarray*}
\hir\tau_0\cdot\tau_0 \!\!\!\! &=& \!\!\!\! 2m^2(\tau_2-\tau_0^2) \\
\hir\tau_1\cdot\tau_1 \!\!\!\! &=& \!\!\!\! 2m^2(\tau_2-\tau_1^2) \\
\hir\tau_2\cdot\tau_2 \!\!\!\! &=& \!\!\!\!
4m^2(\tau_0\tau_1\tau_3-\tau_2^2) \\
\hir\tau_j\cdot\tau_j \!\!\!\! &=& \!\!\!\! 4m^2(\tau_{j-1}\tau_{j+1}-
\tau_j^2)\ \ (3\leq j\leq n-3) \\
\hir\tau_{n-2}\cdot\tau_{n-2} \!\!\!\! &=& \!\!\!\! 4m^2(\tau_n
\tau_{n-1} \tau_{n-3} - \tau_{n-2}^2) \\
\hir\tau_{n-1}\cdot\tau_{n-1} \!\!\!\! &=& \!\!\!\! 2m^2(\tau_{n-2} -
\tau_{n-1}^2) \\
\hir\tau_n\cdot\tau_n \!\!\!\! &=& \!\!\!\! 2m^2(\tau_{n-2}-\tau_n^2).
\end{eqnarray*}
For $\la=2$ it is found that
\medbreak
$$\de_0^{(1)}=-\de_1^{(1)}=1,\ \de_j^{(1)}=0,\ \de_j^{(2)}=(-1)^j\
(2\leq j\leq n-2)$$
\noindent and
$$\de_{n-1}^{(1)}=-\de_n^{(1)}=\pm 1\ (\hbox{n even}),\
\de_{n-1}^{(1)} = -\de_n^{(1)} =\pm i\ (\hbox{n odd}).$$

For $\la=\la_a\ (1\leq a\leq n-2)$, whether $n$ is even or odd,
\medbreak
$$\de_0^{(1)}=\de_1^{(1)}=1,\ \de_{n-1}^{(1)}=\de_n^{(1)} = (-1)^a $$
\noindent and
$$\de_j^{(1)}={2\cos((2j-1)\vth_a)\over \cos\vth_a},\ \de_j^{(2)}=1\
(2\leq j\leq n-2).$$
\bigbreak
\subsection{The $e_6^{(1)}$ theory}
\bigbreak
The Dynkin diagram for $e_6^{(1)}$ is shown in Figure 3.3a.

\centerline{
\begin{picture}(180,165)(0,-20)
\put ( 18,50){\line( 1, 0){24}}
\put ( 58,50){\line( 1, 0){24}}
\put ( 98,50){\line( 1, 0){24}}
\put (138,50){\line( 1, 0){24}}
\put ( 90,58){\line( 0, 1){24}}
\put ( 90,98){\line( 0, 1){24}}
\put ( 10,50){\circle{10}}
\put ( 50,50){\circle{10}}
\put ( 90,50){\circle{10}}
\put (130,50){\circle{10}}
\put (170,50){\circle{10}}
\put ( 90,90){\circle{10}}
\put ( 90,130){\circle{10}}
\put (10,35){\makebox(0,0){$\alpha_1$}}
\put (90,35){\makebox(0,0){$\alpha_4$}}
\put (50,35){\makebox(0,0){$\alpha_3$}}
\put (130,35){\makebox(0,0){$\alpha_{5}$}}
\put (170,35){\makebox(0,0){$\alpha_{6}$}}
\put (105,130){\makebox(0,0){$\alpha_0$}}
\put (105, 90){\makebox(0,0){$\alpha_2$}}
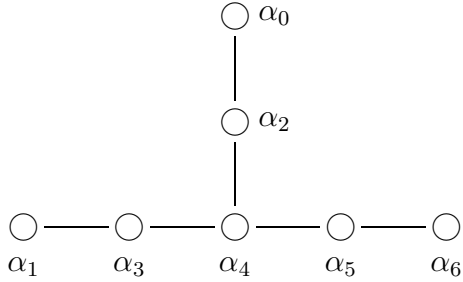
\put (90,10) {\makebox(0,0){Figure 3.3a: Affine Dynkin diagram for
$e_6^{(1)}$.}}
\end{picture}
}
\medbreak
The eigenvalues of the matrix $NC$ are given by

$$\la_1=\la_6=3-\sqrt 3,\ \la_2=2(3-\sqrt 3)$$
\noindent and
$$\la_3=\la_5=3+\sqrt 3,\ \la_4=2(3+\sqrt 3).$$
As in the other simply-laced cases $\eta_j=1$ and $p_j=n_j \ \forall
j$ giving the equations of motion
\begin{eqnarray*}
\hir\tau_a\cdot\tau_a \!\!\!\! &=& \!\!\!\! 2m^2(\tau_b-\tau_a^2) \\
\hir\tau_b\cdot\tau_b \!\!\!\! &=& \!\!\!\! 4m^2(\tau_a\tau_4-\tau_b^2) \\
\hir\tau_4\cdot\tau_4 \!\!\!\! &=& \!\!\!\! 6m^2(\tau_2\tau_3\tau_5-\tau_4^2)
\end{eqnarray*}
\noindent where (a,b)=(0,2),(1,3) and (6,5). A summary of the
$\de$-values for the six single soliton solutions is given in Table 3.3.
\medbreak
With reference to Table 3.3, the vector $\delta^{(1)}=(\delta_0^{(1)},
\delta_1^{(1)},...,\delta_6^{(1)})^{T}$
is an eigenvector of the matrix $K$, which is conjugate to $NC$. The
terms $\de_a^{(b)}\ (b\geq 2)$ are coefficients of $e^{b\Phi}$ in
$\tau_a$. As usual,the $\de$-values corresponding to $\la=0$ have not been
included as they lead to a trivial solution.
\medbreak
\pagebreak
\centerline{
\begin{tabular}{||l|cccccc||}
\multicolumn{7}{l}{Table 3.3: $\delta$-values for $e_6^{(1)}$}\\ \hline
$\lambda$ & $3-\surd 3$ & $3-\surd 3$ & $2(3+\surd 3)$ & $2(3-\surd
3)$ & $3+ \surd 3$ & $3+\surd 3$ \\ \hline
$\delta_0^{(1)}$ & 1 & 1 & 1 & 1 & 1 & 1 \\
$\delta_1^{(1)}$ & $\omega$ & $\omega^2$ & 1 & 1 & $\omega$ &
$\omega^2$\\ $\delta_2^{(1)}$ & $-(\lambda-2)$ & $-(\lambda-2)$ &
$-(\lambda-2)$ & $-(\lambda-2)$ & $-(\lambda-2)$ & $-(\lambda-2)$ \\
$\delta_2^{(2)}$ & 1 & 1 & 1 & 1 & 1 & 1 \\
$\delta_3^{(1)}$ & $-\omega (\lambda-2)$ & $-\omega^2 (\lambda-2)$ &
$-(\lambda-2)$ & $-(\lambda-2)$ & $-\omega (\lambda-2)$ & $-\omega^2
(\lambda-2)$\\
$\delta_3^{(2)}$ & $\omega^2$ & $\omega$ & 1 & 1 & $\omega^2$ & $\omega$\\
$\delta_4^{(1)}$ & 0 & 0 & $3(\lambda-3)$ & $3(\lambda-3)$ & 0 & 0 \\
$\delta_4^{(2)}$ & 0 & 0 & $3(\lambda-3)$ & $3(\lambda-3)$ & 0 & 0 \\
$\delta_4^{(3)}$ & 1 & 1 & 1 & 1 & 1 & 1 \\
$\delta_5^{(1)}$ & $-\omega^2 (\lambda-2)$ & $-\omega (\lambda-2)$ &
$-(\lambda-2)$ & $-(\lambda-2)$ & $-\omega^2 (\lambda-2)$ & $-\omega
(\lambda-2)$\\
$\delta_5^{(2)}$ & $\omega$ & $\omega^2$ & 1 & 1 & $\omega$ & $\omega^2$\\
$\delta_6^{(1)}$ & $\omega^2$ & $\omega$ & 1 & 1 & $\omega^2$ &
$\omega$\\ \hline
\end{tabular}
}
\medbreak
\bigbreak
%
%
\subsection{The $e_7^{(1)}$ theory}
\bigbreak
\centerline{
\begin{picture}(260,100)(0,0)
\put ( 18,50){\line( 1, 0){24}}
\put ( 58,50){\line( 1, 0){24}}
\put ( 98,50){\line( 1, 0){24}}
\put (138,50){\line( 1, 0){24}}
\put (178,50){\line( 1, 0){24}}
\put (218,50){\line( 1, 0){24}}
\put (130,58){\line( 0, 1){24}}
\put ( 10,50){\circle{10}}
\put ( 50,50){\circle{10}}
\put ( 90,50){\circle{10}}
\put (130,50){\circle{10}}
\put (170,50){\circle{10}}
\put (210,50){\circle{10}}
\put (250,50){\circle{10}}
\put (130,90){\circle{10}}
\put (10,35){\makebox(0,0){$\alpha_0$}}
\put (90,35){\makebox(0,0){$\alpha_3$}}
\put (50,35){\makebox(0,0){$\alpha_1$}}
\put (130,35){\makebox(0,0){$\alpha_4$}}
\put (170,35){\makebox(0,0){$\alpha_5$}}
\put (210,35){\makebox(0,0){$\alpha_6$}}
\put (250,35){\makebox(0,0){$\alpha_7$}}
\put (145, 90){\makebox(0,0){$\alpha_2$}}
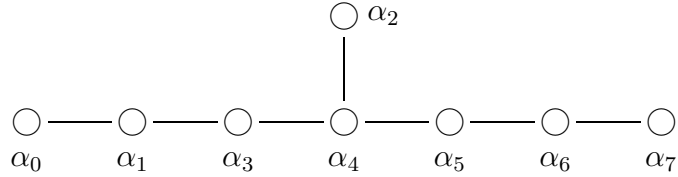
\put (130,10) {\makebox(0,0){Figure 3.4a: Affine Dynkin diagram for
$e_7^{(1)}$.}}
\end{picture}
}
For the $e_7^{(1)}$ theory, whose Dynkin diagram is shown in Figure
3.4a, the non-zero eigenvalues of the matrix $NC$ are
$$\la_1=8\sqrt 3\sin\left ({\pi\over 18}\right )\sin\left ({2\pi\over
9}\right )\ \ \ \ \ \ \ $$
$$\la_2=8\sin^2\left ({2\pi\over 9}\right)\ ,\ \la_3=8\sin^2\left
({\pi\over 3}\right )\ \ \ \ \ \ \ \ \ $$
$$\la_4=8\sqrt 3\sin\left({7\pi\over 18}\right )\sin\left({4\pi\over
9}\right)\ ,\ \la_5=8\sin^2\left ({4\pi\over 9}\right )$$
$$\la_6=8\sqrt 3\sin\left ({5\pi\over 18}\right)\sin\left ({\pi\over
9}\right )\ ,\ \la_7=8\sin^2\left ({\pi\over 9}\right ).$$
\medbreak
A summary of the $\de$-values for the seven soliton solutions for
$e_7^{(1)}$ is given in Table 3.4.
\medbreak
\centerline{
\begin{tabular}{||l|ccc||}
\multicolumn{4}{l}{Table 3.4: $\delta$-values for $e_7^{(1)}$}\\ \hline
$\lambda$ & $\lambda_3$ & $\lambda_2$,$\lambda_5$,$\lambda_7$ &
$\lambda_1$,$\lambda_4$,$\lambda_6$\\ \hline
$\delta_0^{(1)}$ & $1$  & 1 & 1 \\
$\delta_1^{(1)}$ & $-4$ & $-(\lambda-2)$ & $-(\lambda-2)$ \\
$\delta_1^{(2)}$ & $1  $& 1 & 1 \\
$\delta_2^{(1)}$ & $-4 $& 0 & $2(\lambda-2)$ \\
$\delta_2^{(2)}$ & $1  $&$-1$ & 1 \\
$\delta_3^{(1)}$ & $3  $&${1\over 2}(\lambda^2-6\lambda+6)$ &${1\over
2}(\lambda^2-6\lambda+6)$\\
$\delta_3^{(2)}$ & $3  $& ${1\over 2}(\lambda^2-6\lambda+6)$ &${1\over
2}(\lambda^2-6\lambda+6)$ \\
$\delta_3^{(3)}$ & $1  $& 1 & 1 \\
$\delta_4^{(1)}$ & $4  $& 0 &$-(\lambda^2-6\lambda+8)$\\
$\delta_4^{(2)}$ & $6  $&$2(\lambda-1)$&$2(2\lambda^2-9\lambda+9)$\\
$\delta_4^{(3)}$ & $4  $& 0 &$-(\lambda^2-6\lambda+8)$\\
$\delta_4^{(4)}$ & $1  $& 1 & 1 \\
$\delta_5^{(1)}$ & $3  $&$-{1\over 2}(\lambda^2-6\lambda+6)$ &${1\over
2}(\lambda^2-6\lambda+6)$\\
$\delta_5^{(2)}$ & $3  $&${1\over 2}(\lambda^2-6\lambda+6)$ &${1\over
2}(\lambda^2-6\lambda+6)$\\
$\delta_5^{(3)}$ & $1  $&$-1$ & 1 \\
$\delta_6^{(1)}$ & $-4 $&$(\lambda-2)$ & $-(\lambda-2)$ \\
$\delta_6^{(2)}$ & $1  $& 1 & 1 \\
$\delta_7^{(1)}$ & $1  $&$-1$ & 1 \\ \hline
\end{tabular}
}
\medbreak
\vspace{0.3in}
In general, when solving the equations of motion, the $\de$'s are
found to be polynomials in the eigenvalues. However, for $e_7^{(1)}$
the calculation can be simplified by using the characteristic
polynomial of $NC$, which for $\la=\la_2,\
\la_5,\ \la_7$, gives
$$\la^3-12\la^2+36\la-24=0$$
\noindent and for $\la=\la_1,\ \la_4,\ \la_6$, gives
$$\la^3-18\la^2+72\la-72=0.$$
\noindent Therefore, the $\de$'s can be written as quadratic or linear
polynomials in $\la$.
\medbreak
%
%
\pagebreak
\subsection{The $e_8^{(1)}$ theory}
\bigbreak
The Dynkin diagram for $e_8^{(1)}$ is shown in Figure 3.5a.
\medbreak
\centerline{
\begin{picture}(280,110)(0,-5)
\put ( 18,50){\line( 1, 0){24}}
\put ( 58,50){\line( 1, 0){24}}
\put ( 98,50){\line( 1, 0){24}}
\put (138,50){\line( 1, 0){24}}
\put (178,50){\line( 1, 0){24}}
\put (218,50){\line( 1, 0){24}}
\put (258,50){\line( 1, 0){24}}
\put ( 90,58){\line( 0, 1){24}}
\put ( 10,50){\circle{10}}
\put ( 50,50){\circle{10}}
\put ( 90,50){\circle{10}}
\put (130,50){\circle{10}}
\put (170,50){\circle{10}}
\put (210,50){\circle{10}}
\put (250,50){\circle{10}}
\put (290,50){\circle{10}}
\put ( 90,90){\circle{10}}
\put ( 10,35){\makebox(0,0){$\alpha_1$}}
\put ( 90,35){\makebox(0,0){$\alpha_4$}}
\put ( 50,35){\makebox(0,0){$\alpha_3$}}
\put (130,35){\makebox(0,0){$\alpha_5$}}
\put (170,35){\makebox(0,0){$\alpha_6$}}
\put (210,35){\makebox(0,0){$\alpha_7$}}
\put (250,35){\makebox(0,0){$\alpha_8$}}
\put (290,35){\makebox(0,0){$\alpha_0$}}
\put (105,90){\makebox(0,0){$\alpha_2$}}
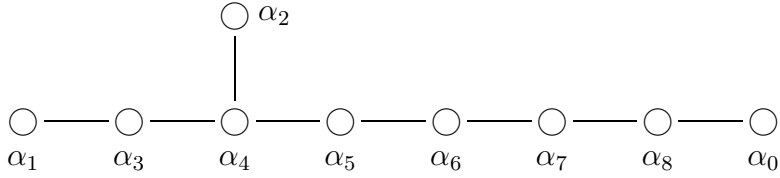
\put (150,10) {\makebox(0,0){Figure 3.5a: Affine Dynkin diagram for
$e_8^{(1)}$.}}
\end{picture}
}
The eigenvalues of the matrix $NC$ are
\begin{eqnarray*}
\la_1\!\!\!\! &=&\!\!\!\! 32\sqrt 3\sin\left ({\pi\over 30}\right
)\sin\left ({\pi\over 5}\right)\cos^2\left ({\pi\over 5}\right ) \\
\la_2\!\!\!\! &=&\!\!\!\! 32\sqrt 3\sin\left ({\pi\over 30}\right
)\sin\left ({\pi\over 5}\right)\cos^2\left ({\pi\over
5}\right)\cos^2\left ({7\pi\over 30}\right ) \\
\la_3\!\!\!\! &=&\!\!\!\! 8\sqrt 3\sin\left ({7\pi\over 30}\right
)\sin\left ({2\pi\over 5}\right) \\
\la_4\!\!\!\! &=&\!\!\!\! 512\sqrt 3\sin\left ({\pi\over 30}\right
)\sin\left ({\pi\over 5}\right)\cos^2\left ({2\pi\over
15}\right)\cos^4\left ({\pi\over 5}\right )  \\
\la_5\!\!\!\! &=&\!\!\!\! 8\sqrt 3\sin\left ({13\pi\over 30}\right
)\sin\left ({2\pi\over 5}\right) \\
\la_6\!\!\!\! &=&\!\!\!\! 8\sqrt 3\sin\left ({11\pi\over 30}\right
)\sin\left ({\pi\over 5}\right) \\
\la_7\!\!\!\! &=&\!\!\!\! 32\sqrt 3\sin\left ({\pi\over 30}\right
)\sin\left ({\pi\over 5}\right)\cos^2\left ({\pi\over 30}\right ) \\
\la_8\!\!\!\! &=&\!\!\!\! 8\sqrt 3\sin\left ({\pi\over 30}\right
)\sin\left ({\pi\over 5}\right).
\end{eqnarray*}
\medbreak
A summary of the $\de$-values for the eight soliton solutions for
$e_8^{(1)}$
is given in Table 3.5.
As in the previous case the characteristic polynomial of $NC$ can be used to
simplify the expressions for the $\de$-values. For $\la=\la_1,\
\la_2,\ \la_4,\ \la_7,$
$$\la^4-30\la^3+240\la^2-720\la+720=0,$$
and for $\la=\la_3,\ \la_5,\ \la_6,\ \la_8,$
$$\la^4-30\la^3+300\la^2-1080\la+720=0.$$
This factorisation of the characteristic polynomial was noted in \cite{bcds1}.
\medbreak
\centerline{
\begin{tabular}{||l|cc||}
\multicolumn{3}{l}{Table 3.5: $\delta$-values for $e_8^{(1)}$}\\ \hline
$\lambda$ & $\lambda_1$,$\lambda_2$,$\la_4$,$\lambda_7$ &
$\lambda_3$,$\lambda_5$,$\lambda_6$,$\la_8$\\ \hline
$\delta_0^{(1)}$ & 1 & 1 \\
$\delta_1^{(1)}$ & $-{1\over 6}(\la^3-24\la^2+132\la-192)$ &${1\over
3}(\la^3-21\la^2+114\la-84)$ \\
$\delta_1^{(2)}$ & 1 & 1 \\
$\delta_2^{(1)}$ & ${1\over 4}(\la^3-18\la^2+84\la-108)$& ${1\over
4}(\la^3-24\la^2+144\la-108)$ \\
$\delta_2^{(2)}$ & ${1\over 4}(\la^3-18\la^2+84\la-108)$& ${1\over
4}(\la^3-24\la^2+144\la-108)$ \\
$\delta_2^{(3)}$ & 1&1 \\
$\delta_3^{(1)}$ &${1\over 6}(\la^3-6\la^2+24)$ &$-{1\over
6}(5\la^3-102\la^2+540\la-384)$\\
$\delta_3^{(2)}$ & ${2\over 3}(5\la^3-60\la^2+225\lambda-261)$&
$-{2\over 3}(\la^3-24\la^2+135\la-99)$ \\
$\delta_3^{(3)}$ & ${1\over 6}(\la^3-6\la^2+24)$ & $-{1\over 6}(5\la^3
-102\la^2+540\la-384)$\\
$\delta_3^{(4)}$ & 1 & 1 \\
$\delta_4^{(1)}$ & $-{1\over 2}(\la-2)(\la^2-6\la+6)$ &$\lambda^2-9\lambda+6$\\
$\delta_4^{(2)}$
&$64\la^3-668\la^2+2214\la-2325$&$3\la^3-50\lambda^2+234\la
-165$\\
$\delta_4^{(3)}$
&$-(303\la^3-3186\la^2+10614\la-11180)$&$-2(3\la^3-54\la^2
+267\lambda-190)$\\
$\delta_4^{(4)}$ &$64\la^3-668\la^2+2214\la-2325$
&$3\la^3-50\la^2+234\la
-165$ \\
$\delta_4^{(5)}$ &$-{1\over 2}(\la-2)(\la^2-6\la+6)$&$\la^2-9\la+6$ \\
$\delta_4^{(6)}$ & 1 & 1\\
$\delta_5^{(1)}$ &${5\over 12}(\la^3-12\lambda^2+48\lambda-60)$
&${5\over 12}(\la^3-18\lambda^2+84\lambda-60)$\\
$\delta_5^{(2)}$ &${5\over 4}(11\la^3-116\lambda^2+384\lambda-400)$ &${5\over
4}(\la-8)(3\lambda^2-26\lambda+20)$\\
$\delta_5^{(3)}$ &${5\over 4}(11\la^3-116\la^2+384\la-400)$&${5\over
4}(\la-8)(3\la^2-26\la+20)$ \\
$\de_5^{(4)}$&${5\over 12}(\la^3-12\la^2+48\la-60)$&${5\over 12}(\la^3
-18\la^2+84\la-60)$\\
$\de_5^{(5)}$&1&1 \\
$\delta_6^{(1)}$ &$-{1\over 6}(\la^3-12\la^2+36\la-24)$ &$-{1\over
6}(\la^3-12\la^2+36\la-24)$ \\
$\delta_6^{(2)}$ &${1\over 6}(7\la^3-78\la^2+288\la-324)$ &${1\over
6}(7\la^3-108\la^2+468\la-324)$ \\
$\de_6^{(3)}$ &$-{1\over 6}(\la^3-12\la^2+36\la-24)$&$-{1\over
6}(\la^3-12\la^2+36\la-24)$\\
$\de_6^{(4)}$ & 1 & 1\\
$\delta_7^{(1)}$ &${1\over 2}(\la^2-6\la+6)$ & ${1\over 2}(\la^2-6\la+6)$ \\
$\de_7^{(2)}$ & ${1\over 2}(\la^2-6\la+6)$ & ${1\over 2}(\la^2-6\la+6)$\\
$\de_7^{(3)}$ & 1 & 1 \\
$\de_8^{(1)}$ &$-(\la-2)$ &$-(\la-2)$\\
$\de_8^{(2)}$ & 1 & 1 \\ \hline
\end{tabular}
}
\pagebreak
\section{Folding and the non-simply-laced algebras}
With the construction of the soliton solutions in the previous
section, enough information has been gathered to deduce solutions to
the non-simply laced algebras.
\bigbreak
\subsection{From \an\ to \cn.}
\bigbreak
The Dynkin diagram for \cn\ is shown in Figure 4.1a.
\medbreak
\centerline{
\begin{picture}(250,90)(0,-20)
\put ( 18,52){\line( 1, 0){24}}
\put ( 18,48){\line( 1, 0){24}}
\put ( 58,50){\line( 1, 0){24}}
\put ( 98,50){\line( 1, 0){14}}
\put (168,50){\line( 1, 0){24}}
\put (208,52){\line( 1, 0){24}}
\put (208,48){\line( 1, 0){24}}
\put (152,50){\line(-1, 0){14}}
\put ( 33,50){\line(-1, 1){10}}
\put ( 33,50){\line(-1,-1){10}}
\put (217,50){\line( 1, 1){10}}
\put (217,50){\line( 1,-1){10}}
\put ( 10,50){\circle{10}}
\put ( 50,50){\circle*{10}}
\put ( 90,50){\circle*{10}}
\put (160,50){\circle*{10}}
\put (200,50){\circle*{10}}
\put (240,50){\circle{10}}
\put (10,35){\makebox(0,0){$\alpha'_0$}}
\put (50,35){\makebox(0,0){$\alpha'_1$}}
\put (90,35){\makebox(0,0){$\alpha'_2$}}
\put (160,35){\makebox(0,0){$\alpha'_{n-2}$}}
\put (200,35){\makebox(0,0){$\alpha'_{n-1}$}}
\put (240,35){\makebox(0,0){$\alpha'_n$}}
\put (125,50){\makebox(0,0){.....}}
\put (125,10) {\makebox(0,0){Figure 4.1a: Affine Dynkin diagram for \cn.}}
\end{picture}
}
The set of roots $\{\al'_i\}$ is expressible in terms of the roots
$\{\al_i\}$ of $a_{2n-1}^{(1)}$ via
$$\al'_0=\al_0\ ,\ \al'_i={1\over 2}(\al_i+\al_{2n-i})\ \ \ (1\leq
i\leq n-1)\ ,\ \al'_n=\al_n.\eqno(4.1.1)$$
This is the origin of the idea of `folding', discussed in \cite{otfold}: the
diagram for $a_{2n-1}^{(1)}$ has been `folded' using
its symmetry under the reflection $0\mapsto 0,i\mapsto 2n-i$ of
the nodes. Generally, suppose the Dynkin diagram of a simply-laced
algebra has some symmetry. The equations of motion then also have
this symmetry, so that solitons with this symmetry as an initial
condition preserve it as they evolve. Thus a solution for
$a_{2n-1}^{(1)}$ can be written in terms of $\{\al'_i\}$ if $\tau_i=
\tau_{2n-i}$. As will be seen these are solutions for \cn.
\medbreak
For \cn\   the $\eta'_j$'s (all quantities relating to \cn\  will be
denoted by a prime) are given by
$$\eta'_0=\eta'_n=1\qquad \hbox{and} \qquad \eta'_i=2 \ (i\neq 0,n)$$
\noindent so that for the single soliton solution $p'_j=1\ \forall j$.
The equations of motion are then
\begin{eqnarray*}
\hir\tau'_0\cdot\tau'_0 \!\!\!\! &=& \!\!\!\! 2m^2(\tau_1^{\prime 2}-
\tau_0^{\prime 2}) \\
\hir\tau'_j\cdot\tau'_j \!\!\!\! &=& \!\!\!\!
2m^2(\tau'_{j-1}\tau'_{j+1} -\tau_j^{\prime 2})\ \ \ (1\leq j\leq n-1) \\
\hir\tau'_n\cdot\tau'_n \!\!\!\! &=& \!\!\!\! 2m^2(\tau_{n-1}^{\prime
2}-\tau_n^{\prime 2}).
\end{eqnarray*}
\noindent This set of equations is that for $a_{2n-1}^{(1)}$ with
$$\tau'_0=\tau_0 \ ,\ \ \tau'_n=\tau_n,\ \ \hbox{and}\ \tau'_j=\tau_j=
\tau_{2n-j}\ \ (1\leq j\leq n-1)\eqno(4.1.2)$$
and so the solutions to \cn\ are those for $a_{2n-1}^{(1)}$ \ with the
conditions (4.1.2) imposed. This leads to the requirement that
$$\omega_a^{j}=\omega_a^{2n-j}\ \ \hbox{where}\ \ 1\leq j\leq n-1.$$
The only $a$ satisfying this equation is $a=n$, giving $\omega_a=-1$,
i.e.
the only non-trivial soliton of $a_{2n-1}^{(1)}$ surviving the folding
procedure is that corresponding to the $n$-th spot on the Dynkin
diagram
(the trivial solution corresponding to the zeroth spot also survives).
Therefore,
$$\tau'_j=1+(-1)^j e^{\Phi}$$
giving the soliton solution to \cn\ as
$$\phi'=-{1\over i\ga}\left (2\sum_{j=1}^{n-1}\al_j\ln\left ({1+(-1)^j
e^{\Phi}\over 1+e^{\Phi}}\right )+\al_n\ln\left ({1+(-1)^n
e^{\Phi}\over 1+e^{\Phi}}\right )\right ).$$

In fact,  with the identification of roots in (4.1.1),
$\phi'=\phi_{(n)}$ where $\phi_{(n)}$ is a soliton solution for
$a_{2n-1}^{(1)}$. This turns out to be a common feature of solitons to
the non-simply-laced theories - they are equal to a soliton of the
corresponding simply-laced algebra.
\bigbreak
%
%
\subsection{From \dn\ to \bn,\ $a_{2n-1}^{(2)},\ d_{n+1}^{(2)},\
a_{2n}^{(2)},\ $and \gtwo}
\bigbreak
Turning first to the \bn\ theory, which has Dynkin diagram shown in
Figure 4.2a,
the set of roots $\{\al'_i\}$ are expressible in terms of the roots
$\{\al_i\}$ of $d_{n+1}^{(1)}$ via
$$\al'_i=\al_i\ (0\leq i\leq n-1),\ \ \al'_n={1\over 2}(\al_n +
\al_{n+1}).$$

\centerline{
\begin{picture}(250,140)(0,5)
\put ( 45, 85){\line(-1,-1){20}}
\put ( 58, 90){\line( 1, 0){24}}
\put ( 98, 90){\line( 1, 0){14}}
\put (168, 90){\line( 1, 0){24}}
\put (152, 90){\line(-1, 0){14}}
\put (208, 92){\line( 1, 0){24}}
\put (208, 88){\line( 1, 0){24}}
\put ( 45, 95){\line(-1, 1){20}}
\put (223, 90){\line(-1, 1){10}}
\put (223, 90){\line(-1,-1){10}}
\put ( 20, 60){\circle{10}}
\put ( 50, 90){\circle{10}}
\put ( 90, 90){\circle{10}}
\put (160, 90){\circle{10}}
\put (200, 90){\circle{10}}
\put (240, 90){\circle*{10}}
\put ( 20,120){\circle{10}}
\put ( 20,105){\makebox(0,0){$\alpha'_0$}}
\put ( 20, 45){\makebox(0,0){$\alpha'_1$}}
\put ( 50, 75){\makebox(0,0){$\alpha'_2$}}
\put ( 90, 75){\makebox(0,0){$\alpha'_3$}}
\put (160, 75){\makebox(0,0){$\alpha'_{n-2}$}}
\put (200, 75){\makebox(0,0){$\alpha'_{n-1}$}}
\put (240, 75){\makebox(0,0){$\alpha'_n$}}
\put (125, 90){\makebox(0,0){.....}}
\put (125, 20){\makebox(0,0){Figure 4.2a: Affine Dynkin diagram for \bn.}}
\end{picture}
}
\noindent With $\tau'_i=\tau_i$,\ $\tau'_n=\tau_n=\tau_{n+1}$, the
equations of motion for  $d_{n+1}^{(1)}$ reduce to those for \bn\ .
The number of solutions is found to be $n-1$ with eigenvalues of $NC$ equal to
$$\la_a=8\sin^2\left ({a\pi\over 2n}\right )\ \ (1\leq a\leq n-1).$$
In this case all the soltions of $d_{n+1}^{(1)}$ survive except those
corresponding to the Dynkin spots $n$ and $n+1$.

Solutions to theories based on twisted algebras such as
$a_{2n-1}^{(2)}$, shown in Figure 4.2b, need to be handled slightly
differently.
The roots of $a_{2n-1}^{(2)}$ are obtainable from those of
$d_{2n}^{(1)}$.
However, if we apply the previous procedure and identify $\tau$'s in the
equations of motion for $d_{2n}^{(1)}$, they are found to be slightly
different from those of $a_{2n-1}^{(2)}$, in that the coefficient of
$m^2$ differs.
This is because the twisted algebras are
obtained from symmetries of the simply-laced diagrams which involve the
extended root, which is thus rescaled by folding.

\centerline{
\begin{picture}(250,140)(0,5)
\put ( 45, 85){\line(-1,-1){20}}
\put ( 58, 90){\line( 1, 0){24}}
\put ( 98, 90){\line( 1, 0){14}}
\put (168, 90){\line( 1, 0){24}}
\put (152, 90){\line(-1, 0){14}}
\put (208, 92){\line( 1, 0){24}}
\put (208, 88){\line( 1, 0){24}}
\put ( 45, 95){\line(-1, 1){20}}
\put (217, 90){\line( 1, 1){10}}
\put (217, 90){\line( 1,-1){10}}
\put ( 20, 60){\circle*{10}}
\put ( 50, 90){\circle*{10}}
\put ( 90, 90){\circle*{10}}
\put (160, 90){\circle*{10}}
\put (200, 90){\circle*{10}}
\put (240, 90){\circle{10}}
\put ( 20,120){\circle*{10}}
\put ( 20,105){\makebox(0,0){$\alpha'_0$}}
\put ( 20, 45){\makebox(0,0){$\alpha'_1$}}
\put ( 50, 75){\makebox(0,0){$\alpha'_2$}}
\put ( 90, 75){\makebox(0,0){$\alpha'_3$}}
\put (160, 75){\makebox(0,0){$\alpha'_{n-2}$}}
\put (200, 75){\makebox(0,0){$\alpha'_{n-1}$}}
\put (240, 75){\makebox(0,0){$\alpha'_n$}}
\put (125, 90){\makebox(0,0){.....}}
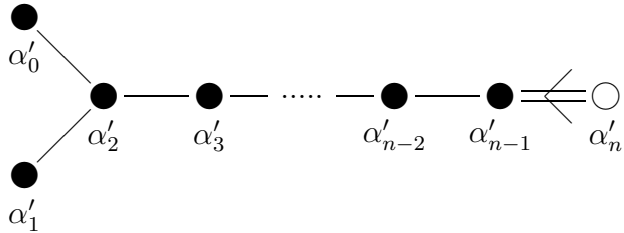
\put (125, 20){\makebox(0,0){Figure 4.2b: Affine Dynkin diagram for
$a_{2n-1}^{(2)}$.}}
\end{picture}
}
The set of roots $\{\al'_i\}$ are expressible in terms of the roots
$\{\al_i\}$ of $d_{2n}^{(1)}$ by
$$\al'_0={1\over 2}(\al_0+\al_{2n-1}),\ \al'_1={1\over
2}(\al_1+\al_{2n}),\ \al'_n=\al_n $$
and
$$\al'_i={1\over 2}(\al_i+\al_{2n-i})\ \ (2\leq i\leq
n-1).$$
It is necessary, therefore, to consider the equations of motion of
$d_{2n}^{(1)}$ with the following identification of $\tau$-functions:
$$\tau'_0=\tau_0=\tau_{2n-1},\
\tau'_1=\tau_1=\tau_{2n},\ \tau'_n=\tau_n\ \ \hbox{and}\ \
\tau'_i=\tau_i=\tau_{2n-i}\ \ (2\leq i\leq n-1)\eqno(4.2.1)$$.
As a result, solutions of $a_{2n-1}^{(2)}$ are those of
$d_{2n}^{(1)}$ with eigenvalue $\la^{(sl)}$ , satisfying (4.2.1) and
$$\sig^2(1-v^2)={1\over 2}m^2\la^{(sl)}=m^2 \la^{(tw)}.$$
With this root convention $\la^{(sl)}$ = $2\la^{(tw)}$,
 $\la^{(tw)}$ being an eigenvalue of the extended Cartan matrix for
$a_{2n-1}^{(2)}$.

As a result, the case $a_{2n-1}^{(2)}$ has $n$ solutions corresponding to
\medbreak
$$\la^{(tw)}_a=4\sin^2\left ({a\pi\over 2n-1}\right )\ \ (1\leq a\leq n-1)\
\hbox{and}\ \la^{(tw)}_n=1.$$

The solitons of $d_{2n}^{(1)}$ lost through folding are those
corresponding to the $k$-th spot \mbox{($1\leq k\leq 2n-1$, k odd)} and one
of the solitons corresponding to the (2n-1)-th and \mbox{2n-th} spots.
\medbreak
This procedure generalises to the other twisted algebras.
\bigbreak
Solitons for the $d_{n+1}^{(2)}$ and $a_{2n}^{(2)}$ theories are
obtained from the $d_{n+2}^{(1)}$ and $d_{2n+2}^{(1)}$ theories
respectively, whereas \gtwo is obtained from $d_4^{(1)}$. The number
of solitons in each case is $n$, $n$, and $1$, respectively.
\bigbreak
\subsection{The remaining theories: $f_4^{(1)},\ d_4^{(3)}\ $and $
e_6^{(2)}$}
\bigbreak
In a similar manner to the previous two subsections, solitons can be
obtained for $f_4^{(1)}$ and $d_4^{(3)}$ from $e_6^{(1)}$, and
$e_6^{(2)}$ from $e_7^{(1)}$. The number of solitons in each case is
two, two and three, respectively.
%
%
\bigbreak
\section{Soliton masses and topological charge}
\bigbreak
\subsection{Soliton mass}
\bigbreak
In \cite{holl} it was shown that the masses of the \an\ solitons are given by
$$M_a={2mh\over \be^2}\sqrt\la_a,\eqno(4.1)$$
\noindent $M_a$ being the mass of the soliton corresponding to
eigenvalue $\la_a$, and $h$ the Coxeter number defined by
$$h=\sum_{j=0}^n n_j.$$
Since the masses of the fundamental Toda particles equal $\sqrt\la$,
the ratios of the soliton masses are equal to the ratios of the
fundamental particles.
\medbreak
By considering the soliton momentum,
$$M\tilde{\ga}(v)v=-\int_{-\infty}^{\infty}dx \dot{\phi}\cdot\phi' $$
\noindent where $\tilde{\ga}(v)=(1-v^2)^{-{1\over 2}}$ it is
straightforward to confirm (case-by-case) that (4.1) holds for the
solitons of the remaining
simply-laced algebras.
\medbreak
Consider now the solitons belonging to the other algebras. For the
untwisted algebras, as each soliton is also a solution of one of the
simply-laced cases, equation (4.1) holds though with the Coxeter
number equal to that of the simply-laced algebra (it is easily shown
that the Coxeter number of an untwisted non-simply-laced algebra is
equal to the Coxeter number of the algebra from which it is folded).
Hence, (4.1) holds with the mass ratios being those of the fundamental
particles.
By relating a solution of a twisted algebra to a solution of the
corresponding simply-laced algebra, the masses of the twisted solitons
are readily seen to satisy (4.1) also.
\vspace{0.3in}
\subsection{Topological charges}
\bigbreak
The topological charge of a soliton is defined as,

$$t={\ga\over 2\pi}\int_{-\infty}^{\infty} dx\partial_x\phi=
{\ga\over 2\pi}(\lim
_{x\rightarrow\infty}-\lim_{x\rightarrow-\infty})\phi(x,t).$$

Previous work \cite{holl} has shown that for \an\ the topological charge of
the soliton $\phi_{(a)}$ is found to be a weight of the $a$-th
fundamental representation. Different choices of $Im\,\xi$ give rise to
different topological charges. For the representations associated with
the roots $\al_1$ and $\al_n$ all weights occur as topological charges
whereas for the other representations a lesser number are found.
We shall use Dynkin labelling for representations, based on the diagrams
in figures 3.1a and 3.2a.

As an example consider the case $a_3^{(1)}$. There are found to be
three non-trivial solutions
\begin{eqnarray*}
\phi_{(1)}\!\!\!\! &=& \!\!\!\!-{1\over i\ga}\sum_{k=1}^3
\alpha_j\ln\left ({1+i^j e^{\Phi} \over 1+e^{\Phi}}\right ) \\
\phi_{(2)}\!\!\!\! &=& \!\!\!\!-{1\over i\ga}\sum_{k=1}^3
\alpha_j\ln\left ({1+(-1)^j e^{\Phi} \over 1+e^{\Phi}}\right ) \\
\phi_{(3)}\!\!\!\! &=& \!\!\!\!-{1\over i\ga}\sum_{k=1}^3
\alpha_j\ln\left ({1+(-i)^j e^{\Phi} \over 1+e^{\Phi}}\right ).
\end{eqnarray*}
\medbreak
The solitons $\phi_{(1)}$ and $\phi_{(3)}$ have topological charges
filling the first and third fundamental representations respectively,
whereas the topological charges
of $\phi_{(2)}$ occur as only some of the weights of the second
fundamental representation.
It may be thought that all the single static solitons have not yet
been found and that other solutions should exist having topological
charges filling the rest of the second fundamental representation.
This does not appear to be true, since it is possible to construct
static double solitons which have
some of the
other weights as topological charges. The static double soliton for
$a_3^{(1)}$ made up of
$\phi_{(1)}$ and $\phi_{(3)}$ has been studied and is found to have
topological charges filling the adjoint representation (1,0,1).
Some consideration has also been given to the
$\phi_{(1)}$-$\phi_{(2)}$ static double soliton which has topological
charges, not previously found, occuring as weights of the
second fundamental representation. However, a study of this case is
not yet complete.
\medbreak
Similar consideration has been given to $d_4^{(1)}$. The solutions
of section 3.2 corresponding to $\la=2$ are found to have topological
charges lying in the first, third and fourth
fundamental representations. The solution corresponding to
$\la=6$ has topological charge lying in the second fundamental
representation.

The static double soliton made up of the pair of solutions associated
with the first and third fundamental representations are found to
have topological charges in the representation (1,0,1,0), as
well as filling up the remainder of the fourth fundamental
representation. Similar results hold for the static
double solitons composed of the other two pairs of solitons having $\la=2$.
\medbreak
It is clear that this aspect of the solitons requires a great deal more study.
\bigbreak
%
%
\pagebreak
\centerline{\Large\bf Acknowledgements}
\medbreak
We should like to thank the U.K. Science and Engineering Research
Council for Studentships, and Ed Corrigan for discussions. N.J.M. wishes
to thank the Japan Society for the Promotion of Science for a
fellowship, and W.A.M. would like to thank both the Department of
Applied
Mathematics and Theoretical Physics, and the Isaac Newton Institute
for
Mathematical Sciences, Cambridge, for their hospitality during the
writing of this paper. W.A.M. wishes also to thank C. A. Hemingway for her
continuing support and encouragement.
\bigbreak
%
%

\end{document}